\documentclass[final,3p,times,twocolumn]{elsarticle}
\usepackage{amssymb}
\usepackage{xcolor}
\usepackage{amsmath}
\journal{Elsevier}

\begin{document}

\begin{frontmatter}

%% Title, authors and addresses

%% use the tnoteref command within \title for footnotes;
%% use the tnotetext command for theassociated footnote;
%% use the fnref command within \author or \address for footnotes;
%% use the fntext command for theassociated footnote;
%% use the corref command within \author for corresponding author footnotes;
%% use the cortext command for theassociated footnote;
%% use the ead command for the email address,
%% and the form \ead[url] for the home page:
%% \title{Title\tnoteref{label1}}
%% \tnotetext[label1]{}

\title{Field-theoretical approach to estimate mean gap and gap distribution in randomly
rough surface contact mechanics}

\author{Yunong Zhou\fnref{add1}\corref{cor1}}\ead{yunong.zhou@yzu.edu.cn}
\author{Hengxu Song\fnref{add2,add3}}
\author{Zhichao Zhang\fnref{add1}}
\author{Yang Xu\fnref{add4,add5}\corref{cor2}}\ead{yang.xu@hfut.edu.cn}
\cortext[cor1]{Corresponding author}
\cortext[cor2]{Corresponding author}
\affiliation[add1]{organization={Department of Civil Engineering},
            addressline={Yangzhou University},
            city={Yangzhou},
            postcode={225127},
            state={Jiangsu},
            country={China}}
\affiliation[add2]{organization={LNM, Institute of Mechanics},
            addressline={Chinese Academy of Sciences},
            city={Beijing},
            postcode={100190},
            country={China}}
\affiliation[add3]{organization={University of Chinese Academy of Sciences},
            city={Beijing},
            postcode={100049},
            country={China}}
\affiliation[add4]{organization={School of Mechanical Engineering},
            addressline={Hefei University of Technology},
            city={Hefei},
            postcode={230009},
            state={Anhui},
            country={China}}
\affiliation[add5]{organization={Anhui Province Key Laboratory of Digital Design and Manufacturing},
            city={Hefei},
            postcode={230009},
            state={Anhui},
            country={China}}

\begin{abstract}
We extend the statistical field-theoretical framework of rough surface
contact mechanics to characterize the interfacial gap between an elastic
half-space and a randomly rough surface incorporating exponential repulsion.
Building upon a cumulant expansion to second-order, we derive an explicit analytical relation
between the mean gap and the applied normal pressure.
This result provides a closed-form expression for the drift and diffusion
coefficients in a convection–diffusion equation governing the scale-dependent
evolution of the gap distribution.
Solving this equation with appropriate initial and boundary conditions yields
the gap distribution under varying external pressures.
Both the mean gap and the gap distribution are found to be in good agreement
with Green’s function molecular dynamics (GFMD) simulations.
Our results demonstrate that the field-theoretical approach enables quantitative predictions not
only for contact stresses but also for interfacial gap in rough surface contacts.
\end{abstract}

%%Graphical abstract
% \begin{graphicalabstract}
% \includegraphics[width=2.0\linewidth]{fig00-graph-abstract}
% \end{graphicalabstract}

%%Research highlights
% \begin{highlights}
% 
% %
% \item We extend field approach to predict gap distribution in rough contact;
% %
% \item We derived explicit pressure-gap relation from the field approach;
% %
% \item Field approach shows good agreement with GFMD simulations;
% 
% \end{highlights}

\begin{keyword}
%% keywords here, in the form: keyword \sep keyword
Contact mechanics \sep interfacial gap \sep field approach \sep rough surface

%% PACS codes here, in the form: \PACS code \sep code

%% MSC codes here, in the form: \MSC code \sep code
%% or \MSC[2008] code \sep code (2000 is the default)

\end{keyword}

\end{frontmatter}

%% \linenumbers

%
\section{Introduction}
\label{sec:intro}

When two nominally flat but microscopically rough surface are pressed together,
only a small portion of the nominal contact area comes into solid-solid contact,
while most of the interface still remains separated.
In such a case, accurate characterization of interfacial properties, such as
real contact area, contact stress distribution and gap distribution is crucial
for predicting performance in applications involving sealing, lubrication, and
thermal or electrical contact
resistance~\cite{Persson2004JPCM,Wang2019AMI,Zhai2017JEM,Jackson2015JAP}.

Over past decades, various theoretical models have been proposed to address
the complexities of rough surface contact~\cite{Greenwood1966RSPA,Bush1975W,
Ciavarella2000RSPA,Persson2001JCP,Shen2023IJMS,Wang2021JT,Ta2021TI,Liang2024TI}.
Among these studies, Persson's theory of contact has gained increasing attention
due to its ability to provide a comprehensive framework that captures
multi-scale roughness and considers the long range elastic
coupling~\cite{Persson2001JCP,Xu2025Arxiv}.
Owing to these advantages, Persson theory has been shown to yield predictions
that are in close agreement with both experiment observations and numerical
simulations~\cite{Persson2007PRL,Bottiglione2009TI,Lyashenko2013TL,Fischer2020TL}.

The central idea of Persson theory is to observe how statistical properties in rough surface
contacts, such as contact stress distribution, evolve with geometrical scale changes.
This process is commonly implemented by starting from an ideally smooth
interface and then gradually adding roughness components corresponding to
smaller wavelengths, which is often described in terms of
increasing ``magnification''~\cite{Prodanov2013TL,Dapp2014JPCM}.

The evolution of contact stress with increasing magnification can also be
modeled as Markov process, in which the evolution of stress distribution
only depends on this most ``recent'' magnification~\cite{Xu2024IJSS}.
In such a way, the Chapman-Kolmogorov equation can be formulated and by
further applying the no re-entry assumption, the diffusion equation that
governing the evolution of stress distribution can be deduced, which finally
leads us to the solution of Persson theory~\cite{Xu2024IJSS,Xu2024TL}.
Building upon this, a wide range of extensions have been developed
to incorporate additional physical effects, including elastic–plastic
deformation, electrostatic adhesion, wear, and percolation
estimation~\cite{Persson2001PRL,Lambert2025PRE,Persson2018JCP,Persson2025JCP,
Dapp2012PRL,Mueller2023PRL}.

Complementary to Persson's development, M{\"u}ser formulated a statistical
field-theoretical framework for contact mechanics, in which the pressure
distribution is obtained through a statistical cumulant expansion, which
reduces to Persson’s theory at the leading order but allows higher-order
corrections to be incorporated.
This formalism not only captures the elastic coupling between asperities but
also provides a consistent method to include non-Gaussian tails in the
pressure distribution when higher-order moments are considered.
Using this formulation, the contact stress distribution has been calculated
to high accuracy for elastic contact with exponential repulsive contacts
~\cite{Mueser2008PRL,Campana2008JPCM}.
In fact, by following the basic idea of statistical field-theoretical approach,
the explicit expression of mean gap can be derived accordingly.

On the other hand, Xu {\it et al.}~\cite{Xu2024JMPS} demonstrated that the evolution of gap
distribution with increasing magnification can be described by the convection-diffusion equation
associated with proper initial and boundary conditions.
Zhou {\it et al.}~\cite{Zhou2026TI} derived the same governing equation for gap distribution by
following the basic idea of Persson's original theory for contact stress
distribution~\cite{Persson2001JCP}.
This differential equation can be solved numerically once the drift and diffusion coefficients,
which can be expressed as the change of mean and variance of the interfacial gap separately,
are specified.
Previous studies demonstrated that the mean gap can be derived based on
Persson's theory directly or indirectly, and shown good agreement with
numerical simulations with proper fudge factor to correct the elastic strain
energy for partial contacts~\cite{Yang2008PRL,Yang2008JPCM,Almqvist2011JMPS}.

In this study, we will extend the field-theoretical approach to address
mean gap in elastic contact between half-space and a rigid, randomly rough
surface incorporating the exponential repulsive interaction.
We will derive the explicit relation between the applied pressure and the
mean gap so that the drift and diffusion coefficient can be determined
accordingly.
Based on this point, solving the convection–diffusion equation with
appropriate initial and boundary conditions yields the gap distribution under
specified applied pressure and magnification.

The theoretical predictions then will be validated with Green's function
molecular dynamics (GFMD) simulations~\cite{Campana2006PRB,Zhou2019PRB,
Venugopalan2017MSMSE} across a range of Hurst exponents and applied pressure.
GFMD is a boundary element method (BEM) that propagate the displacement modes of surface
layer in Fourier space according to Newton's equation of motion until the local minimum of
total potential is achieved.
It has been widely used in elastic contact simulations incorporating multi-scale roughness and
various boundary conditions~\cite{Zhou2022TL,Zhou2024JT,vanDokkum2018TL}.

The remainder of this paper is organized as follows:
The model investigated in this study are introduced in Sect.~\ref{sec:model}.
The derivation of mean gap and related results and discussions are presented in
Sect.~\ref{sec:mean-gap}.
The related results on gap distribution are summarized in
Sect.~\ref{sec:gap-dist} and conclusions would be sketched in
Sect.~\ref{sec:conclusion}.

\section{Model}
\label{sec:model}

\begin{figure*}[ht]
\begin{center}
\includegraphics[width=0.9\linewidth]{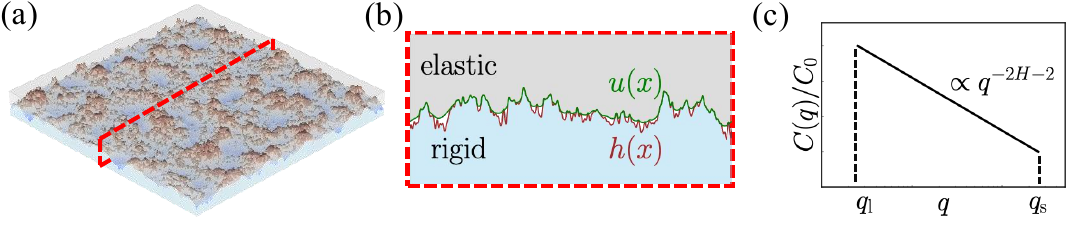}
\caption{
\label{fig:schematic}
(a) Schematic of frictionless contact between elastic half-space and
rigid randomly rough surface.
(b) Cross-section of deformed elastic half-space and rough surface;
(c) A typical power spectral density (PSD) $C(q)$ of randomly rough surface.
}
\end{center}
\end{figure*}

As shown in Fig.~\ref{fig:schematic}, we consider the frictionless
contact between an elastic half-space and a rigid randomly rough surface.
The elastic half-space is originally flat, linearly elastic with plane strain
modulus $E^* \equiv E / (1 - \nu^2)$, where $E$ denotes the elastic modulus
and $\nu$ the Poisson ratio.
The external pressure $\sigma_0$ is uniformly distributed at the far end of the elastic
half-space, which is equivalent to the case that $\sigma_0$ applied at the center-of-mass mode
of the elastic half-space in Fourier space.

The randomly rough surface is fixed in space, and the height spectrum obeys the
random phase approximation, that is, the Fourier coefficient of height
$\tilde h({\mathbf q}) = \sqrt{C(q)/A}\exp(i 2\pi X_q)$, where $X_q$ denotes
a pseudo random number generator that yields values uniformly distributed
within the interval $[0, 1]$, and $A$ denotes the nominal contact area which is fixed to
unity by default.
The height power spectrum $C(q)$, which is presented in
Fig.~\ref{fig:schematic}(c), is defined as
\begin{equation}
\label{eq:spectrum}
C(q) = C_0 \left(\frac{q}{q_{\mathrm l}}\right)^{-2H-2},
\end{equation}
where $H$ denotes the Hurst exponent, $C_0$ the pre-factor that can be fully determined as
long as either the root-mean-square height or root-mean-square gradient is specified,
$q = \vert {\mathbf q} \vert = \zeta q_{\mathrm l}$ denotes the magnitude of
the wave vector $\mathbf q$ that refined in the interval $[q_{\mathrm l}, q_{\mathrm s}]$,
where $q_{\mathrm l} = 2\pi/\lambda_{\mathrm l}$ denotes the longest wave number,
$q_{\mathrm s} = 2\pi/\lambda_{\mathrm s}$ the shortest cut-off and $\zeta$ the magnification.

Therefore, the height of the periodically repeated rough surface can be expressed as the
inverse Fourier transform
\begin{equation}
  h({\mathbf r}) = \frac{A}{(2\pi)^2} \int {\mathrm d}^2q~
  \tilde h({\mathbf q})\exp(i{\mathbf q}\cdot {\mathbf r}).
\end{equation}
The associated Fourier transform then can be written as
\begin{equation}
  \tilde h({\mathbf q}) = \frac{1}{A} \int {\mathrm d}^2r~
  h({\mathbf r})\exp(-i{\mathbf q}\cdot {\mathbf r}).
\end{equation}

The interaction between counter faces is defined by an exponentially repulsive
potential function, which is given as
\begin{equation}
  U_{\mathrm{int}} = \gamma_0 \int{\mathrm d}^2r~e^{-g({\mathbf{r}})/\rho},
\end{equation}
where $\gamma_0$ denotes the surface energy, $\rho$ represents the interaction
range and $g({\mathbf{r}})$ the interfacial gap, which is given by
\begin{equation}
  g({\mathbf{r}}) = g_0 + u({\mathbf{r}}) - h({\mathbf{r}}),
\end{equation}
where $g_0$ represents the mean gap, $u({\mathbf{r}})$ denotes the surface
displacement of elastic half-space with $\langle u({\mathbf{r}}) \rangle = 0$
and $h({\mathbf{r}})$ the height of randomly rough surface with
$\langle h({\mathbf{r}}) \rangle = 0$, which can be fully determined by the
inverse Fourier transform of $\tilde h({\mathbf q})$.

The elastic energy stored in the deformed counterface can be written as
\begin{equation}
U_{\mathrm{ela}} = \frac{E^*A}{4}\sum_{{\mathbf q}}q\vert \tilde u({\mathbf q})  \vert^2,
\end{equation}
where $\tilde u({\mathbf q})$ denotes the Fourier coefficient of surface displacement of elastic
manifold.
The work of external pressure can be given as
\begin{equation}
U_{\mathrm{ext}} = \sigma_0 A g_0.
\end{equation}
Therefore, the total potential is
\begin{equation}
\begin{aligned}
  U_{\mathrm{tot}} &= \frac{E^*A}{4}\sum_{{\mathbf q}}q\vert \tilde u({\mathbf q}) \vert^2 +
  \gamma_0 \int{\mathrm d}^2r~e^{-g({\mathbf{r}})/\rho} \\
  &+ \sigma_0 A g_0.
\end{aligned}
\end{equation}

\section{Mean gap}
\label{sec:mean-gap}

\begin{figure*}[hbpt]
\begin{center}
\includegraphics[width=0.95\linewidth]{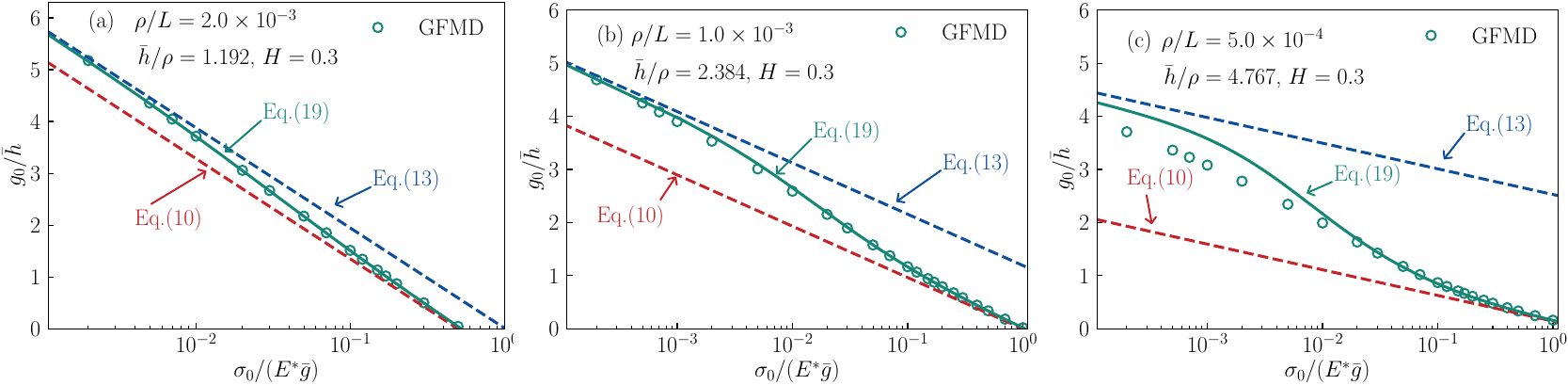}
\caption{
Reduced mean gap $g_0/\bar h$ as a function of reduced external
pressure $\sigma_0/(E^*\bar g)$ for interaction range $\rho/L$
equals to (a) $2.0 \times 10^{-3}$, (b) $1.0 \times 10^{-3}$ and (c) $5.0 \times 10^{-4}$
with Hurst exponent $H = 0.3$,
where $L$ denotes the system size and $\bar g$ the root-mean-square gradient of height.
Red dashed line represents the large pressure limit, blue dashed line denotes the small pressure
limit, the solid line denotes the field approach and open circles the GFMD simulation results.
Specifically, the longest wavelength $\lambda_{\mathrm l} / L = 0.25$, the shortest cut-off
$\lambda_{\mathrm s} / L = 2.5 \times 10^{-3}$ and the root-mean-square gradient of surface height
$\bar g$ is fixed to unity.
The surface energy $\gamma_0 / (E^*L) = 1.0 \times 10^{-3}$.
}
\label{fig:meangap-load-H03}
\end{center}
\end{figure*}

The relation between the mean gap $g_0$ and the applied pressure $\sigma_0$
follows from the equilibrium condition
$\partial U_{\mathrm{tot}} / \partial g_0 = 0$.
Evaluating this condition for the total energy, we obtain
\begin{equation}
  \label{eq:mean-gap-exact}
  g_0(\sigma_0) = \rho \ln\left\{ \frac{\gamma_0}{\rho A \sigma_0}
  \int {\mathrm d}^2r~e^{[h({\mathbf r}) - u({\mathbf r})]/\rho}\right\}.
\end{equation}
This expression is exact within the present exponential repulsive interaction
model.

Two limiting cases can be easily identified.
First, in the limit of $\sigma_0 \gg E^* \bar g$, where $\bar g$ denotes the root-mean-square
gradient of height and is set to unity if not mentioned explicitly, the deformed surface of the
elastic half-space conforms the rigid rough surface, so
that $u({\mathbf r}) \approx h({\mathbf r})$.
In this case the integral reduces to the nominal contact area, and one obtains
\begin{equation}
\label{eq:right-limit}
  g_0(\sigma_0 \gg E^*\bar g)\approx\rho\ln\left(\frac{\gamma_0}{\rho\sigma_0}\right).
\end{equation}
Second, in the opposite limit where $\sigma_0 \ll E^* \bar g$, the elastic displacements are
negligible, that is $u({\mathbf r}) \approx 0$, and thus
\begin{equation}
  g_0(\sigma_0 \ll E^* \bar g) = \rho\ln\left\{ \frac{\gamma_0}{\rho A \sigma_0}
\int {\mathrm d}^2r~e^{h({\mathbf r})/\rho} \right\}.
\end{equation}
The remaining integral contains the statistical contribution of the roughness
profile $h({\mathbf r})$.
Since $h({\mathbf r})$ is assumed to follow Gaussian distribution with zero mean,
we can write the cumulant expansion as
\begin{equation}
  \ln\langle e^{h({\mathbf r})/\rho}\rangle = \sum_{n=1}^{\infty}\frac{1}{\rho^n n!}\kappa_n,
\end{equation}
where $\langle \cdots \rangle$ represents the ensemble averaging and $\kappa_n$ denotes the $n$-th
order cumulant of $h({\mathbf r})$.
Since $h({\mathbf r})$ is assumed to follow Gaussian distribution with zero
mean, all cumulants of order higher than two vanish, i.e.,
$\kappa_1 = \langle h({\mathbf r}) \rangle = 0$, $\kappa_2 = {\mathrm{Var}}[h({\mathbf r})]$ and
$\kappa_{n \ge 3} = 0$.
Therefore, we get
\begin{equation}
\label{eq:left-limit}
\begin{aligned}
  g_0(\sigma_0 \ll E^*\bar g) &= \rho \ln \left\{ \frac{\gamma_0}{\rho \sigma_0}
  e^{{\mathrm{Var}}[h({\mathbf r})]/(2\rho^2)}\right\} \\
&= \rho\ln\left(\frac{\gamma_0}{\rho\sigma_0}\right) + \frac{\bar h^2}{2\rho},
\end{aligned}
\end{equation}
where $\bar h$ represents the root-mean-square height of the rough surface.
Thus, in the low-pressure regime the mean gap receives an additive correction
proportional to the mean-square height of rough surface.

Between these two asymptotic limits lies the crossover regime
$\sigma_0 \simeq E^* \bar g$, where the competition between elastic energy and
repulsive interaction determines the interfacial gap.
To describe this situation, we should start from Eq.~(\ref{eq:mean-gap-exact}) by expanding
the effective field $h({\mathbf r}) - u({\mathbf r})$ to second-order cumulant expansion,
and using the Parseval's theorem, we get
\begin{equation}
\label{eq:mean-gap-approx}
\begin{aligned}
g_0(\sigma_0) &\approx \rho\ln\left\{
\frac{\gamma_0}{\rho\sigma_0}
\exp\left[\sum_{{\mathbf q}} \frac{\vert \tilde h({\mathbf q})-
\tilde u({\mathbf q})\vert^2}{2\rho^2}\right]
\right\} \\
&= \rho\ln\left( \frac{\gamma_0}{\rho\sigma_0} \right) +
  \frac{1}{2\rho}\sum_{\mathbf q} \vert \tilde h({\mathbf q})-
  \tilde u({\mathbf q})\vert^2 \\
&\rightarrow \rho\ln\left( \frac{\gamma_0}{\rho\sigma_0} \right) +
  \frac{A}{4\pi\rho}\int{\mathrm d}q~q\vert \tilde h({\mathbf q})-
  \tilde u({\mathbf q})\vert^2.
\end{aligned}
\end{equation}

In the Fourier space, if the system is well behaved, the elastic
displacement $\tilde u({\mathbf q})$ can be expanded as a power series of
the $\tilde h({\mathbf q})$, which reads
\begin{equation}
\tilde u({\mathbf q}) = G_1({\mathbf q})\tilde h({\mathbf q}) + 
  \sum_{\mathbf q'} G_2({\mathbf q}, {\mathbf q'})\tilde h({\mathbf q}-{\mathbf q'}) \tilde h({\mathbf q'}) + \cdots,
\end{equation} 
where $G_n({\mathbf q}, {\mathbf q'}, \cdots)$ denotes the expansion coefficients and should
depend on the applied pressure and the interaction between counter faces.
To obtain the leading order of this expression, we write the total potential to second-order
cumulant expansion, which reads
\begin{equation}
\begin{aligned}
\frac{U_{\mathrm{tot}}}{A}
  &=
  \frac{E^*}{4} \sum_{{\mathbf q}} q \vert \tilde u({\mathbf q}) \vert^2 + \sigma_0 g_0 \\
  &+ \gamma_0 \exp\left\{ -\frac{g_0}{\rho} +
  \frac{1}{2\rho^2}\sum_{{\mathbf q}} \vert \tilde h({\mathbf q}) - \tilde u({\mathbf q}) \vert^2 \right\}.
\end{aligned}
\end{equation}
Taking the first derivative with respect to $\tilde u({\mathbf q})$, we get
\begin{equation}
\begin{aligned}
\frac{1}{A}\frac{\partial U_{\mathrm{tot}}}{\partial \tilde u({\mathbf q})} &=
\frac{E^*}{2} \sum_{{\mathbf q}} q \tilde u({\mathbf q})
- \frac{\gamma_0}{\rho^2} \sum_{{\mathbf q}} \left[ \tilde h({\mathbf q})
- \tilde u({\mathbf q}) \right] \\
&\times \exp\left\{
-\frac{g_0}{\rho} + \frac{1}{2\rho^2} \sum_{{\mathbf q}} \vert \tilde h({\mathbf q})
- \tilde u({\mathbf q}) \vert^2
\right\} \\
&= \frac{E^*}{2} \sum_{{\mathbf q}} q \tilde u({\mathbf q}) - 
\frac{\sigma_0}{\rho} \sum_{{\mathbf q}} \left[ \tilde h({\mathbf q})
- \tilde u({\mathbf q}) \right].
\end{aligned}
\end{equation}
This derivative requires vanish for arbitrary $\tilde u({\mathbf q})$, therefore,
\begin{equation}
\label{eq:disp-height}
  \tilde u({\mathbf q}) \approx \frac{1}{1 + \eta\zeta} \tilde h({\mathbf q}),
\end{equation}
where we introduced the dimensionless parameter $\eta = \rho q_{\mathrm l}E^*/(2\sigma_0)$.
More details about the derivation can be refered in M\"user's pioneering work~\cite{Mueser2008PRL}.
Substituting Eq.~(\ref{eq:disp-height}) into Eq.~(\ref{eq:mean-gap-approx}),
and using the height spectrum given in Eq.~(\ref{eq:spectrum}) leads to
\begin{equation}
\label{eq:mean-gap-field}
\begin{aligned}
g_0(\sigma_0) &\approx \rho\ln\left( \frac{\gamma_0}{\rho\sigma_0} \right) \\
&+ \frac{H\bar{h}^2}{\rho} \int_1^{\zeta}{\mathrm d}\zeta'
\left( \frac{\eta\zeta'}{1 + \eta\zeta'} \right)^2 \zeta'^{-2H-1}.
\end{aligned}
\end{equation}
The first term is the contribution coming from the homogeneous exponential
repulsion, while the second term encodes the correction due to surface
roughness.
This integral representation smoothly interpolates between the two asymptotic
limits, i.e., for $\eta \ll 1$ ($\sigma_0 \gg E^* \bar g$), it reduces to
Eq.~(\ref{eq:right-limit}) while for $\eta \gg 1$ ($\sigma_0 \ll E^* \bar g$), it
reproduces Eq.~(\ref{eq:left-limit}).

\begin{figure*}[hbpt]
\begin{center}
\includegraphics[width=0.95\linewidth]{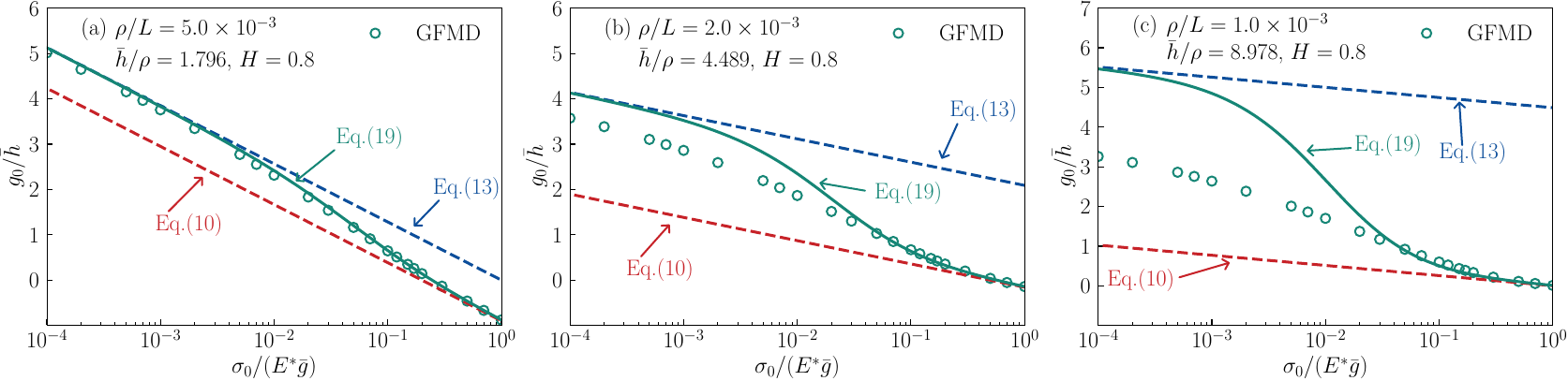}
\caption{
Reduced mean gap $g_0/\bar h$ as a function of reduced external
pressure $\sigma_0/(E^*\bar g)$ for interaction range $\rho/L$
equals to (a) $5.0 \times 10^{-3}$, (b) $2.0 \times 10^{-3}$ and (c) $1.0 \times 10^{-3}$ with
Hurst exponent $H = 0.8$.
Red dashed line represents the large pressure limit, blue dashed line denotes the small pressure
limit, the solid line denotes the field approach and open circles the GFMD simulation results.
Specifically, the longest wavelength $\lambda_{\mathrm l} / L = 0.25$, the shortest cut-off
$\lambda_{\mathrm s} / L = 2.5 \times 10^{-3}$ and the root-mean-square gradient of surface height
$\bar g$ is fixed to unity.
The surface energy $\gamma_0 / (E^*L) = 1.0 \times 10^{-3}$.
}
\label{fig:meangap-load-H08}
\end{center}
\end{figure*}

Fig.~\ref{fig:meangap-load-H03} presents a comparison of predictions of the mean gap $g_0$ as a
function of applied pressure $\sigma_0$ between the field approach and GFMD for various
interaction range $\rho$ at $H = 0.3$.
It is shown that for relatively large ranges of the repulsive potential $\rho$, e.g., in the
cases where $\rho / L = 2.0 \times 10^{-3}$ and $\rho / L = 1.0 \times 10^{-3}$, the theoretical
and numerical results exhibit good agreement, indicating that under conditions where the potential
varies smoothly and the contact response remains approximately linear, the first-order
approximation of the field approach provides a quantitatively reliable description of the mean
gap as a function of applied pressure in rough surface contacts.
However, as shown in Fig.~\ref{fig:meangap-load-H03}(c), for small value of $\rho$, i.e., when the
repulsive potential become short-ranged and steep, the field approach systematically overestimates
the mean gap compared to GFMD results when pressure is small, especially when
$\sigma_0/(E^*\bar g) < 1.0 \times 10^{-2}$.

For a larger Hurst exponent $H = 0.8$, the agreement becomes more restrictive with respect to
the interaction range.
As shown in Fig.~\ref{fig:meangap-load-H08}(a), good quantitative agreement between theory and
GFMD is obtained only for relatively large interaction range $\rho/L = 5 \times 10^{-3}$.
When $\rho / L$ is reduced to smaller values, as shown in Figs.~\ref{fig:meangap-load-H08}(b-c),
significant deviations emerges between the theoretical and numerical results.

The origin of this discrepancy could be explained by the linearization assumption adopted in the
field-theoretical approach, which is based on the first-order approximation of the interfacial
displacement with respect to the surface roughness in Fourier space.
This assumption is valid when the local displacement of elastic surface is much smaller than
the surface roughness and when the repulsive interaction varies smoothly with the interfacial
gap.
However, when $\rho$ becomes small, the gradient of the repulsive potential increases sharply,
giving rise to pronounced nonlinear response, which makes the first-order approximation
fails to capture the nonlinear effect.

In such a case, higher-order cumulant corrections would be required to capture the accurate
behavior in the short-range interaction case.
Unfortunately, this correction is not applicable to the hard-wall constraint.
The reason is that the hard-wall constraint is not a smooth function, which involves a
discontinuous change in displacement as surfaces approach each other.
This lack of smoothness prevents the system from being described by a power series expansion,
which is essential for the higher-order corrections~\cite{Mueser2008PRL}.

\begin{figure}[htbp]
\begin{center}
\includegraphics[width=0.95\linewidth]{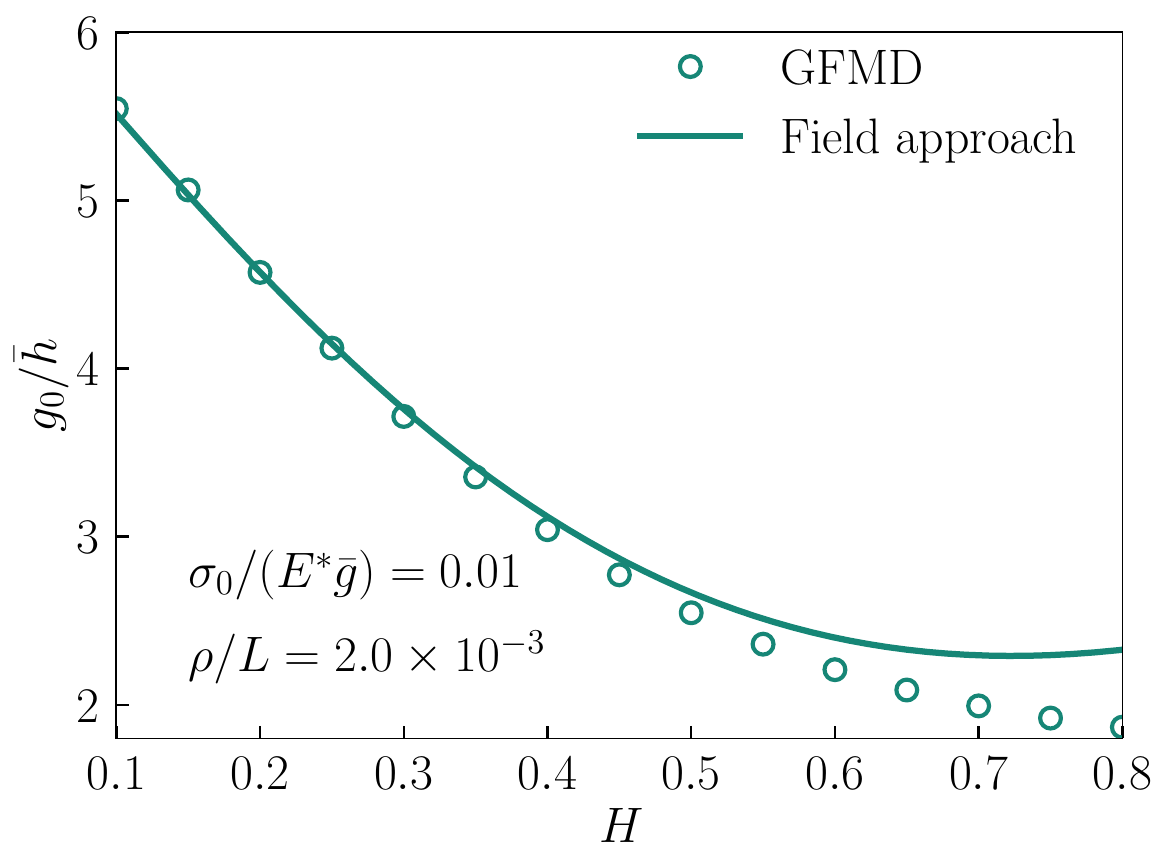}
\caption{
Reduced mean gap $g_0/\bar h$ as a function of Hurst exponent $H$ in the case of
$\rho/L = 2.0\times 10^{-3}$ and $\sigma_0 / (E^* \bar g) = 0.01$.
The solid line denotes the field approach and open circles the GFMD simulation results.
}
\label{fig:meangap-hurst-a}
\end{center}
\end{figure}

\begin{figure}[htbp]
\begin{center}
\includegraphics[width=0.95\linewidth]{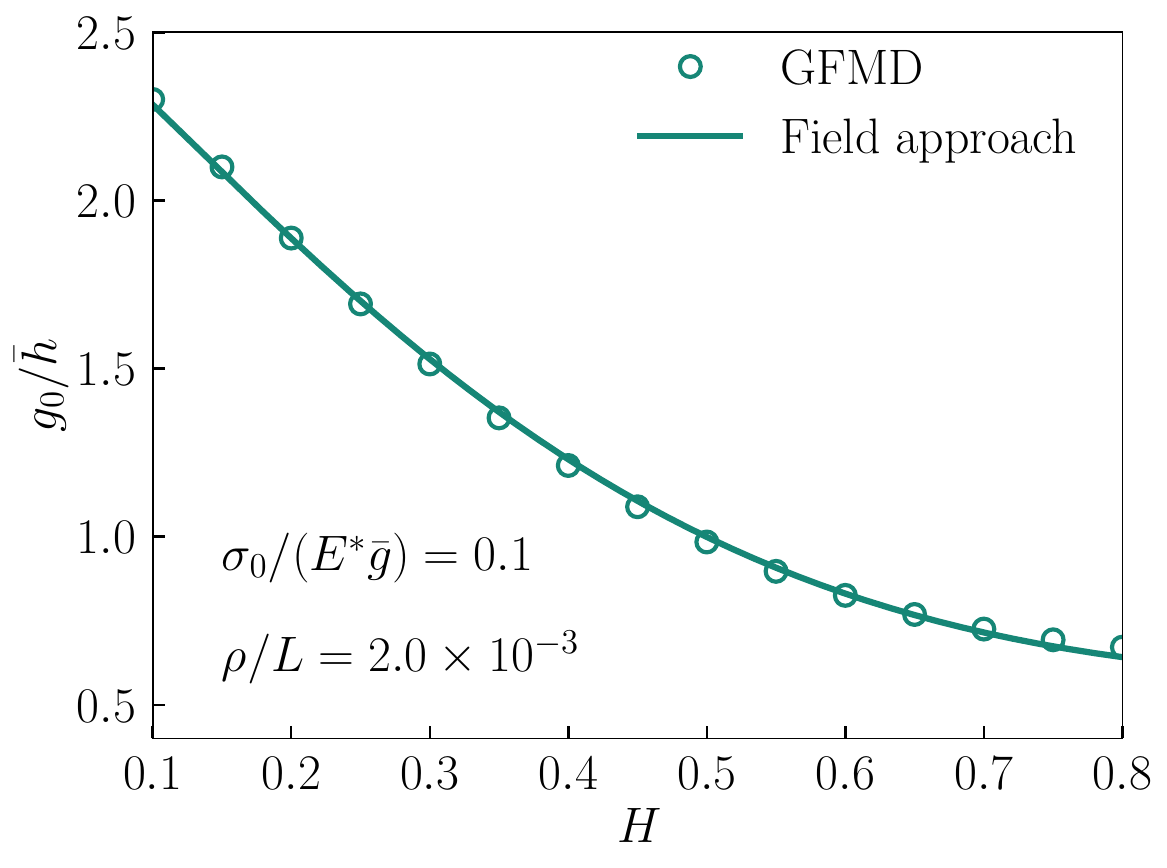}
\caption{
Exactly the same as Fig.~\ref{fig:meangap-hurst-a}, while $\sigma_0 / (E^* \bar g) = 0.1$.
}
\label{fig:meangap-hurst-b}
\end{center}
\end{figure}

On the other hand, it can be observed in
Fig.~\ref{fig:meangap-load-H03} and Fig.~\ref{fig:meangap-load-H08} that with increasing
load $\sigma_0$, the theoretical predictions recover good consistency with GFMD.
This observation is further comfirmed by comparison of Fig.~\ref{fig:meangap-hurst-a} and
Fig.~\ref{fig:meangap-hurst-b}, in which larger external pressure leads to better consistency
between field approach and numerical simulations.
The reason is that with increasing external load, the mean gap decreases and a larger fraction of
asperities come into contact.
This suppresses the influence of local nonlinearity and statistical tail effects in the gap
distribution, effectively restoring the validity of the linear approximation.
Consequently, in the high-load regime the displacement field becomes closer to the surface
roughness, and the theoretical predictions consistent with GFMD, even for surfaces with relatively
large Hurst exponents.

For relatively small external load, as shown in Fig.~\ref{fig:meangap-hurst-a}, small Hurst
exponents ($H < 0.5$) leads to good agreement between field approach and GFMD, while discrepancy
occurs with increasing Hurst exponent to $H > 0.5$.
As suggested by one of the anonymous reviewers, these discrepancies are closely related to the
ratio between root-mean-square height and interaction range $\bar h / \rho$.
This indeed make sense, because for a given moderate applied pressure, when $\bar h$ becomes
larger, the surface's local asperities increases, leading to stronger elastic coupling between
contact regions.
This in turn enhances the non-linearity of contact,making the first-order approximation less
reliable.
On the other hand, when $\rho$ is smaller, the effect of short-range interactions become more
pronounced, and the effects of roughness becomes more significant, further compromising the
accuracy of the first-order approximation.

Additionally, when the root-mean-square gradient $\bar g$ is fixed to unity, it is easy to
demonstrate that the $\bar h$ monotonically increases with increasing Hurst exponent $H$.
Thus, the ratio $\bar h / \rho$ becomes larger as $H$ increases, and leads to greater deviation
between the theoretical approximation and GFMD results, which is observed in
Fig.~\ref{fig:meangap-hurst-a}.

This conclusion is further confirmed in Fig.~\ref{fig:meangap-load-H03} and
Fig.~\ref{fig:meangap-load-H08}.
In both cases of $H = 0.3$ and $H = 0.8$, we observed that as $\bar h/\rho$ increases, the deviation
between the theory and GFMD becomes more pronounced.
This trend further reinforces the idea that surface roughness and interaction range play crucial
roles in the accuracy of the field approach, especially when dealing with higher Hurst exponents
and smaller interaction ranges.

Despite the deviations at higher Hurst exponents, the field approach remains of considerable
theoretical and practical value as it provides an explicit analytical framework that captures the
essential scaling behavior of the mean gap and applied pressure.
Moreover, although many engineering surfaces exhibit Hurst exponents
around $0.8$~\cite{Persson2014TL,Gujrati2018ACSAMI,Gujrati2021STMP}, there also
exist a substantial number of surfaces with $H < 0.5$~\cite{Hinkle2020SA,Monti2022JMPS}, where
the field approach performs particularly well.
Therefore, even though quantitative discrepancies may arise in certain parameter regimes, the
field approach continues to serve as a valuable method to investigate the contact problems between
rough surfaces.

\section{Gap distribution}
\label{sec:gap-dist}

The knowledge of explicit expression of mean gap, as presented in Eq.~(\ref{eq:mean-gap-field}),
allows us to further investigate the gap distribution $P(g, \zeta)$ with specified applied
pressure.
It can be conducted by considering the convection-diffusion equation as follows
\begin{equation}
\label{eq:convection-diffusion}
  \frac{\partial P}{\partial\zeta} = -D_1(\zeta) \frac{\partial P}{\partial g} + D_2(\zeta)
  \frac{\partial^2P}{\partial g^2},
\end{equation}
where $D_1(\zeta)$ denotes the drift coefficient and $D_2(\zeta)$ the diffusion
coefficient~\cite{Xu2024JMPS,Zhou2026TI,Xu2025JT}.
It is equivalent to state that the gap should follow the stochastic differential
equation (SDE)~\cite{Risken1989,Ito1951NMJ}
\begin{equation}
\label{eq:sde}
\frac{{\mathrm d}g}{{\mathrm d}\zeta} = D_1(\zeta) + \Gamma(\zeta),
\end{equation}
where $\Gamma(\zeta)$ denotes the random noise that obey
\begin{equation}
\begin{aligned}
&\langle \Gamma(\zeta) \rangle = 0, \\
&\langle \Gamma(\zeta)~\Gamma(\zeta') \rangle = 2D_2(\zeta)~\delta(\zeta-\zeta'),
\end{aligned}
\end{equation}
where $\langle \cdots \rangle$ denotes the ensemble averaging and $\delta(\cdots)$ represents the
Dirac function.
Eq.~(\ref{eq:sde}) allows us to simulate the gap distribution directly as long as the explicit
expression of $D_1(\zeta)$ and $D_2(\zeta)$ are specified.

The drift term represents the deterministic change of the mean gap induced by the progressive
inclusion of the shorter wavelength components into the roughness, the drift coefficient $D_1(\zeta)$ is given by the
first derivative of the mean gap $g_0$, which is given by Eq.~(\ref{eq:mean-gap-field}), with
respect to the magnification $\zeta$.
The diffusion coefficient $D_2(\zeta)$ denotes the statistical broadening of the gap distribution
due to the random roughness introduced when the magnification $\zeta$ is increased
by an infinitesimal increment, which is defined as
\begin{equation}
  D_2(\zeta) = \frac{1}{2}\lim_{\Delta\zeta\rightarrow 0}
  \frac{\langle [g(\zeta+\Delta\zeta) - g(\zeta)]^2 \rangle}{\Delta\zeta}.
\end{equation}
When the magnification is increased from $\zeta$ to $\zeta + \Delta\zeta$, only surface roughness
with wave-vectors $\vert {\mathbf q} \vert$ in a narrow range
$[\zeta q_{\mathrm l}, (\zeta + \Delta\zeta)q_{\mathrm l}]$ are newly introduced.
The corresponding height increment $\delta h({\mathbf r})$ is therefore a superposition of
statistical independent Fourier components with random phases.

According to the analysis presented above, the incremental elastic displacement can be written
as
\begin{equation}
  \delta \tilde u({\bf q}) \approx \frac{1}{1 + \eta\zeta} \delta \tilde h({\bf q}).
\end{equation}
Consequently, we can write
\begin{equation}
\begin{aligned}
  \langle (\delta g)^2 \rangle &\approx \left( \frac{\eta\zeta}{1 + \eta\zeta} \right)^2
  \langle (\delta h)^2 \rangle \\
  &\propto \left( \frac{\eta\zeta}{1 + \eta\zeta} \right)^2 \zeta^{-2H-1}\Delta\zeta
\end{aligned}
\end{equation}
and as a result, the diffusion coefficient can be obtained as
\begin{equation}
  D_2(\zeta) \approx C_1 \left( \frac{\eta\zeta}{1 + \eta\zeta} \right)^2 \zeta^{-2H-1},
\end{equation}
where $C_1$ denotes a dimensionless pre-factor.

After specifying the drift and diffusion coefficients, a crucial issue in the current stage
concerns the treatment of the boundary conditions as the elastic slab approaches the rough
counter face.
In traditional contact models with hard-wall constraints, the interfacial gap is strictly
non-negative~\cite{Xu2024JMPS}.
In the present work, however, the interaction is assumed as exponential repulsion, which gives
a repulsive force
\begin{equation}
  f_{\mathrm{rep}} = \frac{\gamma_0}{\rho}\exp(-g/\rho).
\end{equation}
This value of force increases rapidly as the gap decreases and therefore provides a strong deterministic
drift that prevents the stochastic particle from penetrating into the negative gap zone.
As a consequence, the gap evolution is governed by a SDE with exponential repulsive drift
near $g = 0$.

In practice, however, the repulsive force entering the SDE should be multiplied by a fudge factor
$C_2$ in order to reproduce the gap distribution obtained from GFMD simulations.
This fudge factor reflects the reduction of the effective local repulsion due to long range
elastic coupling in rough surface contacts, thus can be interpreted as an effective
renormalization of the interaction strength and should be quantitatively smaller than unity.

\begin{figure}[htbp]
\begin{center}
\includegraphics[width=0.9\linewidth]{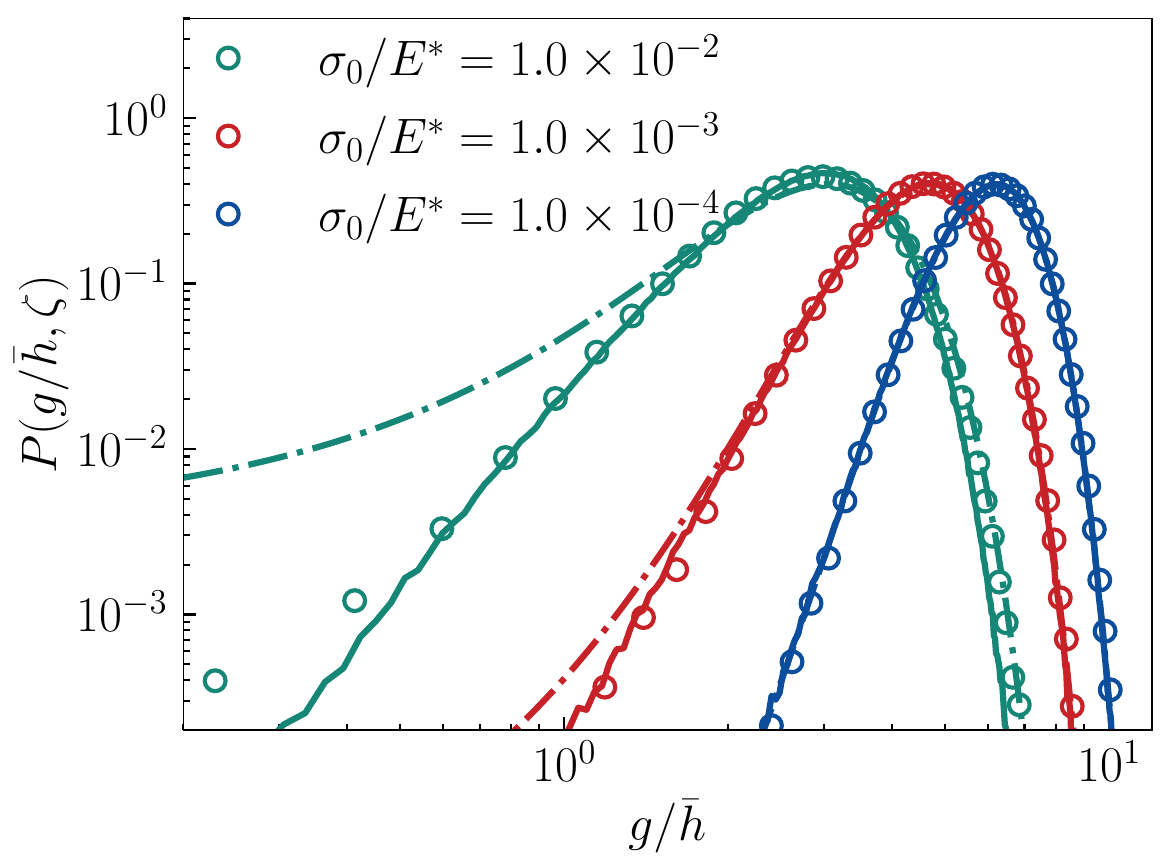}
\caption{
\label{fig:gapDist}
Gap distribution $P(g, \zeta)$ predicted by GFMD (open symbols) and field approach (solid lines)
with varying external pressure.
Dashed lines denote the low-pressure limits that can be calculated according to Eq.~(50) in Ref.\cite{Xu2024JMPS}.
Specifically, the pre-factor $C_0/L^4 = 8.0 \times 10^{-6}$, the Hurst exponent is fixed to
$H = 0.3$, the longest wavelength $\lambda_{\mathrm l}/L = 0.25$, the shortest cut-off
$\lambda_{\mathrm s}/L = 2.5\times 10^{-3}$, the surface
energy $\gamma_0/(E^* L) = 1.0\times 10^{-3}$ and the interaction range
$\rho/L = 2.0 \times 10^{-3}$.
}
\label{fig:gapDist}
\end{center}
\end{figure}

The SDE for the interfacial gap is solved using a Monte Carlo scheme based on the Euler-Maruyama
method.
The drift and diffusion coefficients are pre-computed on the CPU as functions of the magnification
and subsequently used to evolve an ensemble of independent gap measurements on the GPU.
Each measurement is advanced from $\zeta = 1$ to the target magnification by incorporating the
deterministic drift, the exponential repulsion and a Gaussian noise term.
The step size $\Delta\zeta$ was chosen to ensure both numerical stability and convergence.
After performing a convergence test, we found that a step size of $\Delta\zeta = 1.0\times 10^{-2}$
provided a good balance between computational efficiency and accuracy.
Further reduction in the step size did not significantly improve the results, but it did increase
computational cost.
All simulations are performed in parallel on our personal cluster equipped with $64$GB memory and
an NVIDIA RTX 4090D GPU, enabling up to $10^{7}$ stochastic realizations within roughly $20$
seconds.

Fig.~\ref{fig:gapDist} compares the gap distribution $P(g, \zeta)$ for varying external pressures
obtained from GFMD simulations with the predictions of the SDE with field approximations of the
drift and diffusion coefficient.
It is observed that, when the value of applied pressure is relatively small, such as
$\sigma_0/E^* = 1.0 \times 10^{-4}$, the gap distribution (solid blue line)
predicted by the field approach are in excellent agreement with both GFMD (open circles) and
prediction by Persson theory (dahsed blue line).
As the external pressure increases, however, systematic deviations gradually emerge, while the
field approach still reproduces the qualitative trend of the distributions, noticeable
quantitative differences appear, particularly in the small-gap region, and become more pronounced
at higher external pressure.

The observed trend can be understood from the change in the dominant contact mechanisms with
increasing external pressure.
At small external pressures, contact is limited to a few isolated asperities and the real contact
area remains small, so the interfacial gap is mainly controlled by the statistical
properties of surface roughness.
In this regime, the elastic response can be well approximated as linear approximation, so that
the field approach remain valid, leading to excellent agreement with GFMD.

As the external pressure increases, long-range elastic coupling between contact patches becomes
increasingly important, and the deformation field can no longer be regarded as a weakly perturbed
response to surface roughness.
Therefore, the leading-order approximation of the displacement in Fourier space underlying the
field approach becomes insufficient.
As a result, higher-order elastic interactions between asperities, fully captured by GFMD, are
only linearly approximated in the field approach, leading to increasing discrepancies with
increasing external pressures.

\section{Conclusions}
\label{sec:conclusion}

In this study, we have extend the rigorous field approach to describe the mean gap and gap
distribution in the contact between an elastic half-space and a randomly rough surface.
Starting from the explicit expression of the mean gap derived based on the rigorous field approach,
the evolution of the gap distribution is formulated in terms of a convection–diffusion equation,
or equivalently, a stochastic differential equation characterized by scale-dependent drift and
diffusion coefficients.
This approach provides a reduced yet physically transparent description of rough contact, in which
the combined effects of surface roughness, elastic deformation, and exponential interfacial
repulsion are coupled into an effective stochastic process.

The proposed framework enables an efficient Monte Carlo simulation of the gap distribution over a
wide range of external loads.
By performing large-scale GPU-accelerated simulations, we demonstrate that the predicted gap
distributions are in very good agreement with reference results obtained from Green’s function
molecular dynamics (GFMD) and Persson theory.
The agreement holds at small applied pressure, indicating that the statistical features of
rough surface contacts can be accurately captured by the field approach with leading order
expansions.
It is expected that the deviation between our approach and GFMD could be decreased by considering
a few more higher order cumulant expansions.

Considering that our results demonstrate that the proposed framework provides a
convincing and efficient alternative for predicting mean gap and gap distribution in rough contact
problems, future work may focus on incorporating adhesion and thermal effects, thereby extending
the applicability of the approach to a broader range of contact mechanics.

\section*{Acknowledgments}

YZ acknowledges the financial supports from the National Natural Science
Foundation of China (Grant No.12402116) and the Opening fund of State Key Laboratory of Nonlinear
Mechanics (Grant No. LNM202510).
HS acknowledges the financial support from the Strategic Priority Research Program
of the Chinese Academy of Science through grant No. XDB0620101.
YX acknowledges the financial support from the Natural Science Foundation of Anhui
Province (Grant No. 2508085ME101).

\section*{Data Availability}
The data are available from the corresponding author upon reasonable request.

\section*{Declaration}
The author declares that he has no conflict interest.

\section*{Declaration of generative AI and AI-assisted technologies in the manuscript preparation process}
During the preparation of this manuscript, the authors used ChatGPT to assist with language
editing and clarity improvement.
All content was carefully reviewed and revised by the authors, who take full responsibility for
the accuracy, originality, and integrity of the published work.

%% The Appendices part is started with the command \appendix;
%% appendix sections are then done as normal sections
\appendix

\section{Proof of the equivalence between Eq.(~\ref{eq:convection-diffusion}) and Eq.(~\ref{eq:sde})}
We assume that the interfacial gap $g$ between rough surfaces obey the stochastic differential
equation
\begin{equation}
  \frac{{\mathrm d}g}{{\mathrm d}\zeta} = D_1(\zeta) + \Gamma(\zeta)
\end{equation}
If the magnification $\zeta$ increases to $\zeta + \Delta\zeta$, the increment of gap $\Delta g$
can be written as
\begin{equation}
\begin{aligned}
  \Delta g &= \int_{\zeta}^{\zeta+\Delta\zeta} {\mathrm d}s~\left[D_1(s) + \Gamma(s)\right] \\
  &\approx D_1(\zeta)\Delta\zeta + \int_{\zeta}^{\zeta+\Delta\zeta}{\mathrm d}s~\Gamma(s)
\end{aligned}
\end{equation}
as long as $\Delta\zeta$ is small enough.
Therefore, we get
\begin{equation}
\label{eq:moment}
\begin{aligned}
\langle \Delta g \rangle &= D_1(\zeta)\Delta\zeta, \\
\langle \Delta g^2 \rangle &= 2D_2(\zeta)\Delta\zeta.
\end{aligned}
\end{equation}
The gap distribution at magnification $\zeta + \Delta\zeta$ is
\begin{equation}
  \label{eq:gap}
\begin{aligned}
  P(g, \zeta+\Delta\zeta) &= \int{\mathrm d}g'~P(g, \zeta+\Delta\zeta | g', \zeta) P(g', \zeta) \\
  &= \int{\mathrm d}{\Delta g}~P(g; g-\Delta g) P(g-\Delta g, \zeta),
\end{aligned}
\end{equation}
where we assume $g' = g - \Delta g$ and define
$P(g; g - \Delta g) \equiv P(g, \zeta+\Delta\zeta | g-\Delta g, \zeta)$.
We expand $P(g-\Delta g, \zeta)$ with Taylor series to second-order, then we get
\begin{equation}
  \label{eq:taylor}
  P(g-\Delta g, \zeta) \approx P(g, \zeta) - \Delta g \frac{\partial P}{\partial g} +
  \frac 1 2 \Delta g^2 \frac{\partial^2P}{\partial g^2}.
\end{equation}
Substituting Eq.~(\ref{eq:taylor}) into Eq.~(\ref{eq:gap}), we get
\begin{equation}
  \label{eq:taylor-2}
  P(g, \zeta+\Delta\zeta) = P(g, \zeta) - \langle \Delta g \rangle \frac{\partial P}{\partial g}
  + \frac 1 2 \langle \Delta g^2 \rangle \frac{\partial^2 P}{\partial g^2}.
\end{equation}
Substituting Eq.~(\ref{eq:moment}) into Eq.~(\ref{eq:taylor-2}) and assume that $\Delta\zeta$ is
small enough, we get
\begin{equation}
  \frac{\partial P}{\partial\zeta} = -D_1(\zeta)\frac{\partial P}{\partial g} +
  D_2(\zeta)\frac{\partial^2P}{\partial g^2}.
\end{equation}

%% \section{}
%% \label{}

%% If you have bibdatabase file and want bibtex to generate the
%% bibitems, please use
%%
\bibliographystyle{elsarticle-num} 
\bibliography{gapFA}

%% else use the following coding to input the bibitems directly in the
%% TeX file.

\end{document}